\documentclass[aps,prl,twocolumn,superscriptaddress]{revtex4-1}

\usepackage{graphicx}

\newcommand{\lsp}{{\sc Lsp}}

\begin{document}

\title{Simulations of
Magnetic Field Generation in
Unmagnetized Plasmas via Beat Wave Current Drive}

\author{D. R. Welch}
\author{T. C. Genoni}
\author{C. Thoma}
\author{N. Bruner}
\author{D. V. Rose}
\affiliation{Voss Scientific, LLC, Albuquerque, NM 87108}

\author{S. C. Hsu}
\affiliation{Physics Division, Los Alamos National Laboratory, Los Alamos, NM 87545}

\date{\today}

\begin{abstract}
This work describes the scientific basis and associated simulation
results for the magnetization of an unmagnetized plasma via beat
wave current drive.   Two-dimensional electromagnetic
particle-in-cell simulations
have been performed for a variety of angles between the injected
waves to demonstrate beat wave generation in agreement
with theoretical predictions of the beat-wave wave vector and
saturation time, revealing new 2D effects.
The simulations clearly demonstrate electron
acceleration by the beat waves and resultant current drive and
magnetic field generation.  The basic process depends entirely on the angle
between the parent waves and the ratio of the beat-wave phase
velocity to the electron thermal velocity.  The wave to magnetic
energy conversion efficiency of the cases examined is as high as
0.2\%. The technique could enable novel
plasma experiments in which the use of magnetic
coils is infeasible.
\end{abstract}


\maketitle

The nonlinear mixing of electromagnetic (EM) waves in plasmas has been
applied to heating, diagnostics, particle acceleration, ionospheric
plasma modification, and current drive~\cite{STEFAN1989}.
This paper describes the application of beat wave
current drive~\cite{COHEN1984,COHEN1988,ROGERS1992}
to magnetize an initially unmagnetized plasma.
The greatest emphasis of prior beat wave studies
was on accelerating electrons to relativistically high energies
and not on the most effective generation
of current for producing significant magnetic field in an
ambient plasma.  Potential
applications of beat wave magnetic field
generation include controlled fusion concepts such as magneto-inertial
fusion (MIF) \cite{slutz:056303,CHANG2011,HSU2012}, and
laboratory experiments on plasma phenomena of astrophysical
interest, {\em e.g.}, \cite{DRAKE2000}.  For lack of immediate
applications, however, the full
complexity of significant current and
field generation from beat wave current
drive has not been addressed.
While early analyses of beat
wave current drive were essentially 1D~\cite{COHEN1988},
more realistic modeling is now required
with actual exploratory experiments underway
\cite{LIU2011}.  For
significant  current drive, the beat wave phase velocity and electron
thermal velocity must be comparable in order to accelerate a useful
number of electrons, and thus both collisionless (Vlasov) and
collisional modeling are required.  The modeling must also be at
least 2D in order to handle essential experimental input
parameters, such as the shapes, widths,
and injection angle of the overlapping EM waves. It is also required to
obtain the resultant spatial distribution of extended return
currents and the magnetic field.
The 2D modeling also allows more realistic testing of predictions from
1D analysis \cite{COHEN1988}. Using the
Large Scale Plasma (\lsp) code~\cite{welch:072702},  we have performed such
a study using a particular EM wavelength and plasma density range
of relevance to MIF\@.

Remote magnetization has been accomplished previously
by shining a laser onto foil targets with wire loops \cite{DAIDO1986,WOOLSEY2001}.
Beat wave magnetization, however,
offers several potential advantages:
(1)~refraction of the
injected high frequency waves is negligible and
placement of the beat wave interaction region
within the plasma can be precise;
(2)~beat waves are produced in a controllable direction depending on
the angle $\theta$ between injected waves; and (3)~current drive
can be accomplished for thermal plasmas via control of the wave phase
velocity which also depends on $\theta$.
For MIF energy applications requiring reduction of
cross-field energy transport, the Hall parameter (ratio of the electron cyclotron
to momentum transfer frequency)  $\omega_c / \nu_m > 1$.
This ratio can be achieved by seeding modest magnetic fields ($B \sim 1$~T)
in lower density regions of imploding plasmas that get amplified geometrically to $\sim 100$~T
at maximum compression \cite{HSU2012}.

\begin{figure}[!tb]
\includegraphics[scale=0.72,bb=50 15 326 460]{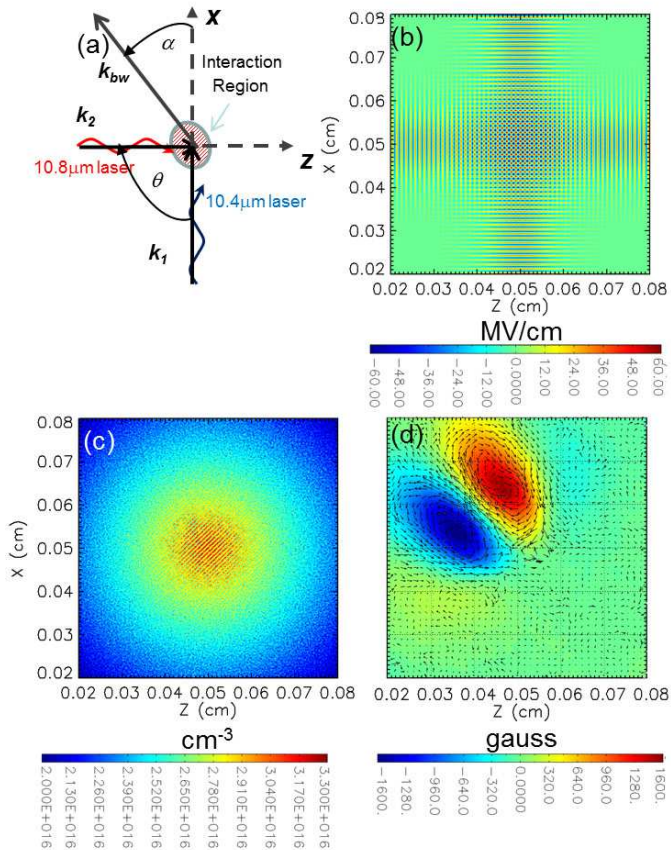}
\caption[Beat wave characteristics.]
{Beat wave magnetic field generation using two lasers of
$3\times 10^{12}$~W/cm$^2$ intensity and
$\theta = 90^{\circ}$ in a 50-eV plasma with $3\times
10^{16}$-cm$^{-3}$ peak density.
(a)~Laser injection and beat wave geometry;
(b)~laser electric fields 20~ps after injection;
(c)~modulated plasma density (at 20~ps); and
(d)~$B$ contours and electron mean velocity vectors (at 100~ps).
\label{fig:beatwave}}
\end{figure}

Figure~\ref{fig:beatwave} shows the setup and representative simulation results
using two CO$_2$ lasers, with intensity of $3\times 10^{12}$
W/cm$^2$ and $\theta=90^\circ$, impinging on
the center of a circular plasma with non-uniform density
peaking at $3\times 10^{16}$~cm$^{-3}$ and initial temperature $T_0$ = 50 eV.
The orientation of the lasers  and the subsequent beat wave are
shown in Fig.~1(a), with the beat wave vector being the difference
between the two laser wave vectors, {\em i.e.}, $\vec{k}_{bw} =
\vec{k}_1 - \vec{k}_2$. The upward- and rightward-propagating lasers have
10.4 and 10.8-$\mu$m wavelengths, respectively.
The combined laser electric field pattern, shown in
Fig.~1(b), drives a plasma beat wave propagating to the upper
left, as shown in Fig.~1(c).  The beat wave modulates the plasma density
by roughly $\pm$10\% with 7-$\mu$m wavelength. The plasma beat wave
accelerates a portion of the plasma electrons to as high as 500~eV in
the $\vec{k}_{bw}$ direction, driving a current and magnetic field structure
shown in Fig.~1(d).

We now discuss the theory of beat wave current drive as exhibited in Fig.~1.
Two waves, with frequencies
$\omega_1$, $\omega_2 \gg$ the electron plasma frequency
$\omega_{pe}$, are launched into a plasma to generate a beat wave with
frequency $\sim \omega_{pe}$ \cite{COHEN1972,ROSENBLUTH1972}.
Resonant interaction between the beat wave and the electron population
is exploited to accelerate electrons via Landau damping and to drive
current and generate magnetic fields.
Consider a beat wave generated from the ponderomotive force ($\vec{E} \times \vec{B}$)
of two intersecting injected waves with electric field $E$ polarization into the
page and wave vector
as shown in Fig.~1(a). The angle $\alpha$ between $\vec{k}_{bw}$ and $\vec{k}_1$
is $\tan(\alpha) = [k_2 \sin(\theta)]/[k_1-k_2\cos(\theta)]$.
If we include the ponderomotive force in the cold electron fluid equation of motion,
the magnitude of the wave vector is
\begin{equation}
k_{bw}  = \frac{k_1-k_2\cos(\theta)}{\cos(\alpha)}.
\label{eq:kbw_mag}
\end{equation}
Furthermore, the rate of increase in the beat wave amplitude is \cite{ROSENBLUTH1972}
\begin{equation}
\dot{A}(t) = \frac{k_{bw} c^{2}}{4\omega_{pe}}\epsilon_{1}\epsilon_{2},
\label{eq:adot}
\end{equation}
where $\epsilon_{i} = eE_{i} / mc\omega_i$, and $\omega_{i}$  is the individual
laser electron oscillation frequency.  For the case that the beat wave
phase velocity $v_{ph} = \omega_{pe}/k_{bw} \ll c$, the wave will break for $A k_{bw} =1$
and saturate \cite{ROSENBLUTH1972}.  Thus, we can estimate the
saturation time of the beat wave from Eqs.~(\ref{eq:kbw_mag}) and
(\ref{eq:adot}) and obtain
\begin{equation}
\tau_{sat} = \frac{1}{\dot{A} k_{bw}}.
\label{eq:tsat}
\end{equation}
Although the beat wave growth rate scales linearly with laser intensity
[$\propto (I_{1} I_{2})^{1/2}$], the wave
saturated amplitude is a weak function of laser intensity. For cases presented in this
paper,
$\tau_{sat} \sim$ 1 ps, providing a lower bound on the simulation duration.

As shown in Fig.~1,  we model the entire beat wave
interaction and current drive using \lsp,
a state-of-the-art, parallel,  multi-dimensional
particle-in-cell (PIC) code.
The simulations resolve the smallest relevant EM wavelengths ($\sim
5$~$\mu$m) and highest frequencies ($\sim 8$~THz). The $> 0.1$-ns
duration and order mm scale length for the interaction demand
$10^5$ time steps and $> 10^6$ cells in 2D\@.
Hundred of particles per cell are required
to adequately resolve the electron energy distribution in the high
energy tail and provide a sufficiently small noise level to discern
the beat wave dynamics.  For these simulations, \lsp~solves the
relativistic Maxwell-Lorentz equations with inter-particle collisions.
Particle scattering is treated
with complete generality via a binary Coulombic interaction algorithm
\cite{Nanbu1998639,welch:072702}.  We use an explicit, energy-conserving
particle advance that sums particle
currents such that charge is conserved~\cite{Welch2001134}.

Our simulation domain
is a 1-mm by 1-mm square enclosing a 0.95-mm diameter plasma cylinder (with vacuum between
plasma edge and domain boundary)  as
shown in Fig.~\ref{fig:beatwave}(c) with density range from 1--$3\times 10^{16}$
cm$^{-3}$.  From the boundaries, we inject 100-$\mu$m
transverse-extent lasers into the plasma. Intensities examined here are relevant for
generating seed magnetic fields of order 5~kG\@.
The resonant plasma density for the case of 10.4 and 10.8~$\mu$m wavelengths is
$1.5 \times
10^{16}$~cm$^{-3}$, half the peak density. We found that
this difference is not significant to the resulting beat
wave production although plasma heating is more efficient
at the resonance. The lasers reach $10^{12}$--$10^{13}$~W/cm$^2$ peak intensity
in 10~ps with $\theta = 20$--180$^{\circ}$.
A range of beat wave characteristics can be obtained by varying $\theta$.
Much past theoretical work
focused on co-linear ($\theta=0$) injection for its relevance to relativistic
electron acceleration. In that case, $v_{ph}$ approaches $c$, which is
suitable for a high energy accelerator, but does not easily couple to a thermal
plasma electron distribution as required for current drive.
Therefore, we focus most of this paper on larger $\theta$.

\begin{figure}
\includegraphics[scale=0.97,bb=4 16 386 490]{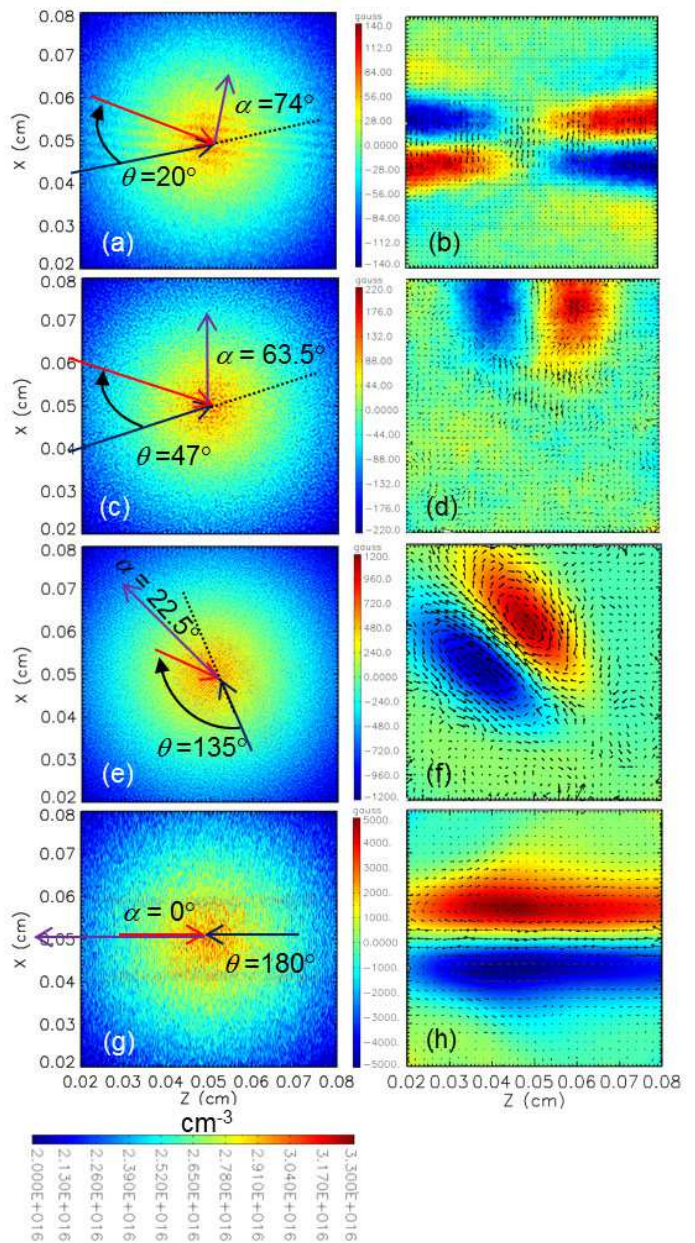}
\caption[Beat wave current drive.]
{Beat wave modulated electron density contours (left column) at $t=20$~ps
for (a)~20$^\circ$, (c)~47$^{\circ}$, (e)~135$^{\circ}$, and (g)~180$^{\circ}$;
magnetic field contours and electron mean velocity vectors (right column)
for (b)~$20^{\circ}$ at 50 ps,
(d)~47$^{\circ}$ at 50 ps, (f) ~135$^{\circ}$ at 100 ps, and (h)~180$^{\circ}$
at 100~ps.  This is for laser intensities of $3\times 10^{12}$~W/cm$^2$ and 50-eV plasma.
\label{fig:magfld}}
\end{figure}

We have verified the simulations against
basic theory. In all but the $\theta = 0$ case, the beat wave amplitude
saturates in several ps, so we present the saturated wave
characteristics after 20~ps in several simulations. The beat wave
characteristics including wavelength and direction are nicely
exhibited by the electron density contours in
Fig.~\ref{fig:magfld} [and Fig.~1(c) for $\theta = 90^{\circ}$]. The simulations agree
closely with the theory for both $\alpha$ and $k_{bw}$.
The smallest $k_{bw}$ is 1/3 that of the lasers
for the $\theta = 20^{\circ}$ case, and the largest, roughly twice that of the lasers,
is found for the
180$^{\circ}$ case. The beat wave saturated amplitude, as measured by
the wave displacement $A$ or electron density modulation,
is predicted to scale inversely with $k_{bw}$.
This is not obvious in the simulations largely
due to the many other competing nonlinear interactions including
temperature dispersion, collisionality, and current drive.  The beat
wave electron density
modulations are all roughly $\pm$10$\%$. Peak beat wave electric field  magnitudes $E_{bw}$
range from 100--1000~kV/cm with only a weak dependence on intensity.

We find that current drive
is higher for larger angles (Fig.~\ref{fig:magfld}). For $T_0$ = 50 eV,
line currents of 8~kA/cm produce $B\sim 2$~kG\@. The $\theta= 20^{\circ}$ and
47$^{\circ}$ simulations show an
order of magnitude smaller fields due to the fast $v_{ph}$ of the beat
waves.  In all cases, the electron current is driven in the direction of
$\vec{k}_{bw}$.  The mean electron velocity vectors in Fig.~\ref{fig:magfld}
reveal a closed current path surrounding
magnetic islands for the $\theta=90$--180$^{\circ}$ simulations. The
lower-$\theta$ simulations have a less obvious return current path due
to the finite volume of the simulation plasma. Higher energy electrons
carrying the current quickly reach the plasma-vacuum interface.

\begin{figure}
\includegraphics[scale=0.45,bb=6 22 411 410]{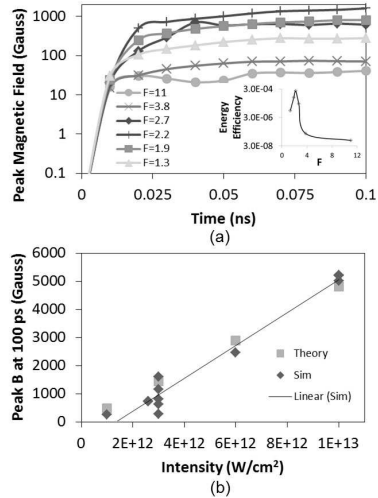}
\caption[Peak magnetic fields.]
{Peak magnetic field driven by beat waves
versus (a)~time at $3\times 10^{12}$~W/cm$^{2}$ and $\theta=90^{\circ}$ for
different values of $F=v_{ph}/v_{te}$,
and (b)~laser intensity for
$\theta \ge 90^{\circ}$ including theoretical values calculated from
Eq.~(\ref{eq:efficiency}) and a linear fit to the simulation data.  The inset in (a) shows the efficiency
of magnetic energy generation versus $F$ at 100~ps.
\label{fig:growthb}}
\end{figure}

Although the beat wave saturation amplitude is only a function of
$\theta$, the beat wave current drive is optimized by controlling
$F=v_{ph}/v_{te}$ (where $v_{te}$ is the electron thermal
velocity), which varies in time by plasma heating
observed at higher laser intensity. In Fig.~\ref{fig:growthb}(a), we show
peak $B(t)$ for several $\theta=90^{\circ}$
simulations with $3 \times 10^{12}$~W/cm$^2$ laser intensity and
$F$ varying from 1.3--11 ($T_0$ = 200--25~eV).  The
simulations show a rapid rise in $B$ within 10~ps
of peak laser intensity at the interaction region. The field
saturation time is consistent with the theoretical value of $\tau_{sat}$. An
optimal value for $F$ is in the range 1.9--2.7
for which $B \sim 1$~kG.

The beat wave current drive efficiency is \cite{COHEN1988}
\begin{equation}
\eta = 0.15 \frac{R_{1}R_{2}}{2 (L_{int}/10~{\rm m})}
\frac{(E_{ph}/10~{\rm keV})^{1.5}}{n_{e} / 10^{13}~{\rm cm}^{-3}} ({\rm A}/{\rm W}),
\label{eq:efficiency}
\end{equation}
where $L_{int}$ is the laser interaction length (roughly 5~cm
in our case).  A
conservative estimate of the action transfers $R_1$ and $R_{2}$
of 0.1 is taken from Fig.~1 of Ref.~3, and
$E_{ph}$ is the electron energy associated with $v_{ph}$.
Using Eq.~(\ref{eq:efficiency}) to estimate peak $B$, we compare
the theory with simulation results as a function of laser
intensity for $\theta \ge 90 ^{\circ}$.  As shown in
Fig.~\ref{fig:growthb}(b), the 1D theory and 2D simulations follow a similar
linear dependence of $B$ on intensity, although there is some variation due
to varying $F$.
We also calculate the total conversion efficiency of laser to magnetic
field energy which we
define as the ratio of magnetic field energy at 100~ps to the total
injected laser energy $E_{inj}$ through 100~ps.
Keeping fixed $\theta = 90 ^{\circ}$
and $3 \times 10^{12}$~W/cm$^2$ intensity ($E_{inj} = 0.015$~J),
we find a maximum in the
energy efficiency of 0.03\% near $F$ = 2.2 with $\eta >
0.001$\% for $F = 1.9$--2.7, as shown in the inset
of Fig.~\ref{fig:growthb}(a). The efficiency falls off steeply for $F
> 3$. The highest efficiency of 0.2\% seen in all our simulations
thus far correspond to $\theta = 180^{\circ}$
and $10^{13}$~W/cm$^2$ ($E_{inj} =0.045$~J)\@.
The efficiency can be optimized via $F$ and maximizing the overlap of injected beams.

 Examining more closely the mechanism of current
 drive in our nominal $3 \times 10^{12}$~W/cm$^2$ intensity and
 $\theta = 90 ^{\circ}$  simulation where $v_{ph} = 0.027 c$, we find
 that beat waves with peak $E_{bw}> 400$~kV/cm can trap electrons
moving in the same direction with sufficient initial speed, such
 that $v > v_{ph} - [4E_{bw}/(mc^{2}k_{bw})]^{1/2} \approx 0.008c$ (16 eV),
 and accelerate them to higher energy. This velocity is determined
 from the total depth of the potential well of the wave in the frame
 of $v_{ph}$.   We injected test electrons in the center of the laser
 interaction region ($X = Z = 0.05$~cm) with ($0.0014$--0.02)$c$ velocity
 in the direction of the beat wave and tracked their velocity ($V^\prime$) and
 position ($X^\prime$) in that direction. Electrons with initial energy $<$ 18 eV
 exhibited oscillatory behavior with peak energy increasing to 50 eV,
 but little net motion. Most higher energy electrons are trapped,
 riding up and down on the beat wave fronts but moving with the beat
 wave. The electrons accelerate to $V^\prime \approx 0.045c$ (500~eV), then down to
 their initial values for several cycles until they are scattered or
 begin to leave the region of laser interaction at $X^\prime >
 0.01$~cm.  At this time, the electrons can retain some
 fraction of their peak accelerated energy. These strongly-directed electrons drive current until
they are scattered or reach the edge of the plasma.
Some of these electrons end up carrying return current as they cycle back.
The detailed relationship between beat wave fields, electron acceleration,
and magnetic field extent will be explored further in a forthcoming paper.

\begin{figure}[!tb]
\includegraphics[width=3.0truein]{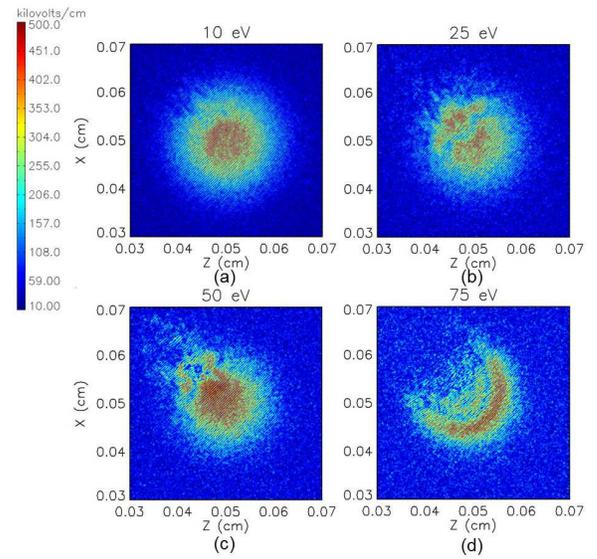}
\caption[Beat wave damping.]
{Electric field magnitude at $t=50$~ps
for $3\times 10^{12}$~W/cm$^2$ laser intensity, $\theta=90^{\circ}$,
and varying $T_0$:  (a)~10~eV ($F = 11$), (b)~25~eV ($F = 3.8$), (c)~50~eV
($F = 2.7$), and (d)~75~eV ($F=2.2$).
\label{fig:damping}}
\end{figure}

The local $E_{bw}$, in  regions of highest electron acceleration and
at optimal $F$, decays due to nonlinear Landau damping. The Landau
damping rate is significant as $k_{bw}\lambda_D \rightarrow 1$.  For
the $F = 2.2$ simulation, $k_{bw}\lambda_D =$ 0.3. The damping effect
on beat wave fields is shown in Fig.~\ref{fig:damping} for varying $T_0$. For this particular beat wave
excitation at a given phase velocity, at low temperature [Fig.~4(a)]
when Landau damping is not strong, the structure reflects
 quite well the beat wave excitation with negligible loss. With higher
temperature [Fig.~4(d)], significant damping is now
evident in the decay of the beat wave excitation magnitude along the
beat wave wavevector direction (at $> 135^{\circ}$ to the $Z$ axis).
Indeed at 200~eV, the beat wave is strongly damped everywhere relative to the
10-eV simulation. It is unclear whether the presence of the magnetic
fields themselves, producing cyclotron radii $\approx 0.02$~cm, also
contributes to the decay of the beat wave. The regions of stronger
beat wave damping also have the highest magnetic field.  The 200-eV
simulation had 1/5 the peak magnetic field but faster damping of the
wave than the 75-eV case. The Landau damping of the beat waves permits
the accelerated electrons to escape the wave structure and drive
current. Thus, the strong wave and current regions remain distinct.

In this paper,
we have described the scientific
basis for magnetization of an unmagnetized plasma
via beat wave current drive.  Specifically,
we performed 2D PIC simulations to explore the scaling of
beat wave production, current drive, and $B$
generation in a non-uniform density plasma
using injected waves near 10-$\mu$m wavelength, corresponding
to CO$_2$ lasers.
Future work should include 3D simulations
and exploration of lower intensity lasers ($10^8$~W/cm$^2$) for
modeling of near term experiments \cite{LIU2011}.
The 2D simulations to 100~ps presented here made use of 64
processors run for roughly 72 hours.  Simulations in 3D will be 100's
of times larger in cell number and require 1000's of processors
on a massively parallel computer.

\begin{acknowledgments}
We acknowledge
excellent code support from R. E. Clark. This work was
supported by the Office of Fusion Energy Sciences of the
U.S. Department of Energy.
\end{acknowledgments}

%

\end{document}